\documentclass[aps,prb,superscriptaddress,twocolumn]{revtex4}
\usepackage{bm,amsmath,graphicx}
\begin{document}                  % DO NOT DELETE THIS LINE

\title{Hybrid reciprocal space for X-ray diffraction in epitaxic layers}

\author{S\'ergio L. Morelh\~ao}
\email[Correspondence e-mail: ]{morelhao@if.usp.br}
\affiliation{Instituto de F\'{\i}sica, Universidade de S\~ao Paulo,
Brazil}
\author{Jarek Z. Domagala}
\affiliation{Institute of Physics, Polish Academy of Sciences,
Warsaw, Poland}

\begin{abstract}
Even after several decades of systematic usage of X-ray diffraction
as one of the major analytical tool for epitaxic layers, the vision
of the reciprocal space of these materials is still a simple
superposition of two reciprocal lattices, one from the substrate
and another from the layer. In this work, the general theory
accounting for hybrid reflections in the reciprocal space of
layer/substrate systems is presented. It allows insight into the
non-trivial geometry of such reciprocal space as well as into many
of its interesting properties. Such properties can be further
exploited even on conventional-source X-ray diffractometers,
leading to alternative, very detailed, and comprehensive analysis
of such materials.
\end{abstract}

\maketitle                        % DO NOT DELETE THIS LINE

\pagebreak

\section{Introduction}

The capability of growing thin layers of single-crystals onto one
face of another single crystal has made possible many fundamental
achievements in semiconductor technology. Epitaxic growth is today
one of the most important and basic processes used in manufacturing
nanostructured devices. Multilayered materials, such as
superlattices and quantum wells, or even quantum wires and dots,
require epitaxy at some stage of their preparation procedures.
X-ray diffraction has been the primary tool for structural analysis
of epitaxic layers, with the associated techniques and machinery
following closely the needs of the semiconductor industry.

A quarter of a century ago, when using a divergent X-ray source and
photographic films to record the layer/substrate diffraction lines
[the simplest possible setup (Chang, 1980) to measure lattice
mismatch of epilayers], Isherwood {\em et al.} (1981) reported the
observation of extra features, a kind of short line, appearing all
over the recorded images. Such features were sequences of
consecutive Bragg reflections in both single-crystal lattices, and
were named hybrid reflections. Latter, the phenomenon was
quantitatively described and methods to exploit its properties were
suggested (Morelh\~ao \& Cardoso, 1991; 1993a; 1993b). However, to
probe the excitement conditions of hybrid reflections precisely,
collimated X-ray beam setups would be necessary, such as those
commonly found in most synchrotron facilities where the beam can be
highly collimated in two orthogonal directions (Morelh\~ao {\em et
al.}, 1991, 1998, 2002, 2003). Such requirements have created a
huge barrier for the systematic usage of this peculiar phenomenon
in the technology of semiconductor devices.

Even after several decades of using X-ray diffraction as one of the
major analytical tools for epitaxic layers, the vision of the
reciprocal space of these materials is still a simple superposition
of two reciprocal lattices: one from the epilayer and the other
from the substrate. Diffraction conditions generating any other
extra feature have been avoided since they could not be explained
within this simplistic vision of the reciprocal space. The
incident-beam optics available for conventional X-ray sources have,
in the past, seemed inappropriate (considering the requirement for
a highly collimated beam in two orthogonal directions) to
investigate azimuthal-dependet features; consequently, the analysis
of epilayers by standard X-ray diffraction techniques would be
compromised when hybrid reflections are excited. This work is an
attempt to change this scenario. Introducing a reciprocal-space
description of hybrid reflections opens the possibility of
exploiting in detail the properties of this phenomenon without the
necessity of using synchrotron facilities. In other words, instead
of avoiding hybrid features when using tube-source diffractometers,
exciting them {\em via} standard reciprocal-space probing
techniques can lead to alternative methods for analyzing epilayered
materials. Here, besides presenting the theory and discussing some
properties of the hybrid reciprocal lattice, experimental examples
are given regarding the type of information that can be accessed by
analyzing the phenomenon properly on commercial diffractometers.

\section{Hybrid reciprocal lattice theory}

Any three-dimensional reciprocal lattice gives rise to a phenomenon
known as $n$-beam diffraction (Colella, 1974; Chang, 1984; Weckert
\& H\"ummer, 1997). Although it can change the relative strength of
Bragg reflections, no extra features are generated in the
reciprocal space since sums of diffraction vectors always end up at
a reciprocal-lattice point (RLP). On the other hand, when two
distinct reciprocal lattices are superposed, as in
epilayer/substrate systems, sum of diffraction vectors may end up
at an empty position of the reciprocal space. This occurs when one
diffraction vector in the sum does not belong to the same lattice
as the others. In this case, hybrid reciprocal-lattice points
(HRLPs) are generated, as illustrated in Fig. 1 and described
mathematically below.

Three-beam X-ray diffractions in single crystals are excited when
the incident beam (wavevector $\bm{k}$ and wavelength $\lambda$),
fulfills two Bragg conditions:
\begin{eqnarray}
\bm{k}\cdot\bm{P} &=& -\bm{P}\cdot\bm{P}/2 {\rm \hspace{.2in} and} \label{eq:1} \\
\bm{k}\cdot\bm{M} &=& -\bm{M}\cdot\bm{M}/2. \label{eq:2}
\end{eqnarray}
Since $\bm{P} = \bm{M} + \bm{N}$, we also have
\begin{equation}
\bm{k}\cdot\bm{N}=-\bm{N}\cdot\bm{N}/2-\bm{N}\cdot\bm{M}
\label{eq:3}
\end{equation}
where $\bm{P}$, $\bm{M}$, and $\bm{N}$ are diffraction vectors of
the primary, secondary and coupling reflections, respectively. The
primary reflection is the one whose intensity is been monitored
while the secondary reflection is brought to diffraction condition
by the crystal azimuthal rotation around $\bm{P}$, as in X-ray
Renninger scanning (Renninger, 1937). Other cases of $n$-beam
diffractions with $n>3$ will be treated here as coincidental
three-beam diffractions.

\begin{figure}
\includegraphics[width=3.2in]{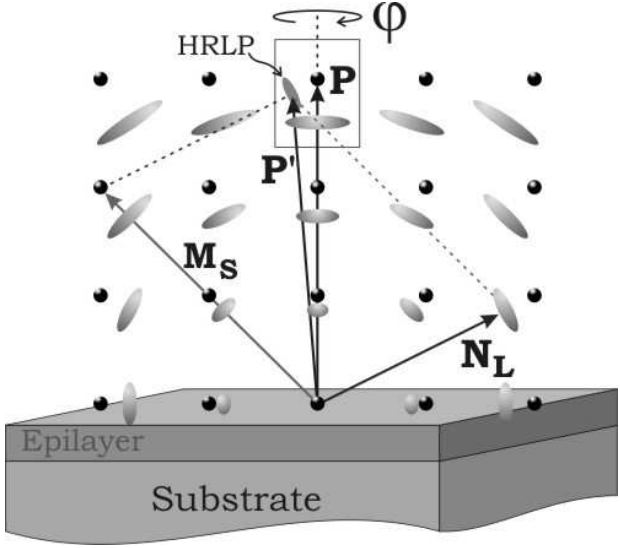}
\caption{In epilayer/substrate systems, superposition of the
substrate (circles) and epilayer (ellipses) reciprocal lattices
gives rise to inter-lattice rescattering processes, as illustrated
by the hybrid reciprocal-lattice point (HRLP) with diffraction
vector ${\nobreak \bm{P}' = \bm{M}_S + \bm{N}_L}$. To excite one
chosen HRLP, the crystal must be at specific azimuthal $\varphi$
angles as detailed in Appendix A.}
\end{figure}

Only equations (\ref{eq:1}) and (\ref{eq:2}) are in fact necessary
to predict three-beam diffractions in a single lattice (Cole {\em
et al.}, 1962; Caticha, 1969), which can be either the epilayer or
the substrate one. However, there are several other similar
diffraction processes: the above-mentioned hybrid reflections,
whose secondary and coupling reflections do not belong to the same
lattice. To predict what should be the exact incident-beam
direction for exciting one of such inter-lattice rescattering
process, equations (\ref{eq:2}) and (\ref{eq:3}) are more suitable
to this purpose, as demonstrated elsewhere for the case of
satellite reflections (Morelh\~ao {\em et al.}, 2003) and
summarized here in Appendix A for the case of epilayer/substrate
systems. Accounting for all possible rescatterings leads to a
reciprocal space that is highly populated with HRLPs and much more
complex than that obtained by just superposing both epilayer and
substrate reciprocal lattices. This hybrid reciprocal space has
been neglected; its features remain unexplored; all knowledge on
this matter is found in reports of a few accidental observations
(Hayashi {\em et al.} 1997, Domagala {\em et al.}, 2006) sometimes
investigated (Morelh\~ao {\em et al.}, 2003), but most of the time
avoided in order not to compromise the system characterization by
standard diffraction techniques, such as rocking curves and
reciprocal-space mapping in triple-axis goniometry.

To visualize the hybrid reciprocal space, let us label the
diffraction vectors of both lattices as
\begin{eqnarray}
\bm{M}_{S,L} &=& h\bm{a}^*_{S,L}+k\bm{b}^*_{S,L}+\ell\bm{c}^*_{S,L} {\rm \hspace{.1in} and} \nonumber \\
\bm{N}_{L,S} &=&
(H-h)\bm{a}^*_{L,S}+(K-k)\bm{b}^*_{L,S}+(L-\ell)\bm{c}^*_{L,S}
\nonumber
\end{eqnarray}
where $S$ and $L$ subscripts stand for the substrate and epilayer
reciprocal-lattice vectors, respectively. $h$, $k$, and $\ell$ are
the Miller indexes of the secondary reflection, and the complete
hybrid reciprocal space around one chosen $HKL$ primary reflection
of the substrate lattice, whose diffraction vector is ${\nobreak
\bm{P}=H\bm{a}^*_S+K\bm{b}^*_S+L\bm{c}^*_S}$, only appears by
rotating the sample around $\bm{P}$ by $360^{\circ}$. The position
of all features regarding the $\bm{P}$ vector is then given by
\begin{equation}
\Delta\bm{P} = \bm{P}' - \bm{P} = h'\Delta\bm{a}^*+
k'\Delta\bm{b}^*+ \ell'\Delta\bm{c}^*, \label{eq:4}
\end{equation}
which is a sub-reciprocal-lattice of points with periodicity
$\Delta\bm{g}^* = \bm{g}^*_L - \bm{g}^*_S$, for $\bm{g}^*$ =
$\bm{a}^*$, $\bm{b}^*$, and $\bm{c}^*$, since $h'$, $k'$, and
$\ell'$ are integer numbers. They stand for either
coupling-reflection ($H-h,K-k,L-\ell$) or secondary-reflection
($h,k,\ell$) indexes according to $\bm{P}' = \bm{M}_S + \bm{N}_L$
or $\bm{P}' = \bm{M}_L + \bm{N}_S$, respectively.

Although equation (\ref{eq:4}) gives the general aspect of the
hybrid reciprocal lattice, there are a few restrictions that should
be considered for each particular system. One is the direction of
the secondary beam (wavevector $\bm{k}_M = \bm{M}_{S,L} + \bm{k}$)
that must cross the epilayer/substrate interface in order for its
respective hybrid diffraction vector, $\bm{P}' = \bm{M}_{S,L} +
\bm{N}_{L,S}$, to be measurable. In other words, if $\hat{\bm{n}}$
is the interface normal direction pointing upwards into the
epilayer and $\xi = \hat{\bm{n}}\cdot \bm{k}_M/|\bm{k}_M|$, we have
that
\begin{equation}
(h',k',\ell') = \left \{
\begin{array}{ccc}
(H-h,K-k,L-\ell) & {\rm if} & \xi > 0 \\
(h,k,\ell) & {\rm if} & \xi < 0.
\end{array}
\right. \label{eq:5}
\end{equation}
Hence $h'k'\ell'$ are the indexes of the Bragg reflection that
occurs in the epilayer lattice. Refraction and total reflection of
the secondary beam, $\bm{k}_M$, at the interface is also a
possibility to be taken into account mainly when $\xi \approx 0$.
In general, epilayers of quaternary alloys grown on miscut
substrates present a relative tilt between both lattices, and this
tilt must be considered when calculating $\Delta\bm{g}^*$.

\subsection{Properties of the hybrid reciprocal lattice}

One of the most interesting properties of the hybrid reciprocal
lattice is that the relative positions $\Delta\bm{P}$ of the HRLPs
depend exclusively on the lattice mismatch between both real
lattices, i.e. their positions do not depend on the X-ray
wavelength. What change with $\lambda$ is the azimuthal angle,
$\varphi$, where each HRLP is excited, as can be calculated by
solving equations (\ref{eq:2}) and (\ref{eq:3}); see Appendix A.
This implies that HRLPs are more easily excited with an X-ray beam
that is poorly collimated in the axial direction, i.e. the
direction perpendicular to the primary incidence plane: the plane
that contains the substrate vector $\bm{P}$, the X-ray source and
the detector system. In commercial high-resolution diffractometers,
the beam is highly conditioned in the incidence plane to about a
few arc seconds, while the axial divergence is of the order of a
few degrees ($\approx 2^o$). On the other hand, in synchrotron
facilities for X-ray diffraction, where the beam is well
conditioned in both directions, the azimuthal positioning of the
sample has to be more accurate if hybrid features are to be
measured (Morelh\~ao {\em et al.}, 2002; 2003).

In Fig.~2, the general properties of hybrid reciprocal lattices are
depicted. Just a few points aligned along the growth direction in
the case of fully strained layers (Fig.~2a), or a well defined
three-dimensional lattice of points around the substrate
reciprocal-lattice points occurs in the case of relaxed layers
(Fig.~2b). Consequently, strain gradients along the layer thickness
give rise to a hybrid lattice of rods instead of points, as
illustrated in Figs.~2(c) and 2(d). Besides the fact that HRLPs are
excited only at certain azimuthal positions, their experimental
observation via conventional reciprocal-space mapping techniques
also require detection optics with some angular acceptance in the
axial direction. When a given HRLP is excited, its diffraction
vector ${\nobreak \bm{P}' = \Delta\bm{P} + \bm{P}}$ is not on the
primary incidence plane and neither is its diffracted beam
${\nobreak \bm{k}' = \bm{P}'+\bm{k}}$. Therefore, visible HRLPs are
those whose diffracted beam, $\bm{k}'$, falls within the angular
range of axial acceptance of the detection system. Moreover, HRLP
positions on reciprocal-space maps correspond to projections of
$\Delta\bm{P}$ on the incidence plane. In terms of longitudinal and
transversal components of the maps, $Q_z$ and $Q_{xy}$
respectively, the HRLPs are seen at
\begin{equation}
Q_z = \Delta\bm{P}\cdot\hat{\bm{z}}~~{\rm and}~~ Q_{xy} =
\Delta\bm{P}\cdot\hat{\bm{k}}_{xy} \label{eq:6}
\end{equation}
where $\hat{\bm{z}} = \bm{P}/|\bm{P}|$ and $\hat{\bm{k}}_{xy} =
[\bm{k}-(\bm{k}\cdot\hat{\bm{z}})\hat{\bm{z}}]/|\bm{k}-(\bm{k}\cdot\hat{\bm{z}})\hat{\bm{z}}|$.
It is then possible to calculate
\begin{equation}
\delta = \lambda Q_{xy}\tan(\Delta\varphi) \label{eq:7}
\end{equation}
as the takeoff angle of $\bm{k}'$ from the incidence plane of the
primary reflection. $\Delta\varphi=\varphi'-\varphi$, and
$\varphi'$ is the azimuthal position of $\bm{P}'$ around $\bm{P}$
with respect to the same reference direction for $\varphi$ and in
the same sense of rotation.

\begin{figure}
\includegraphics[width=3.2in]{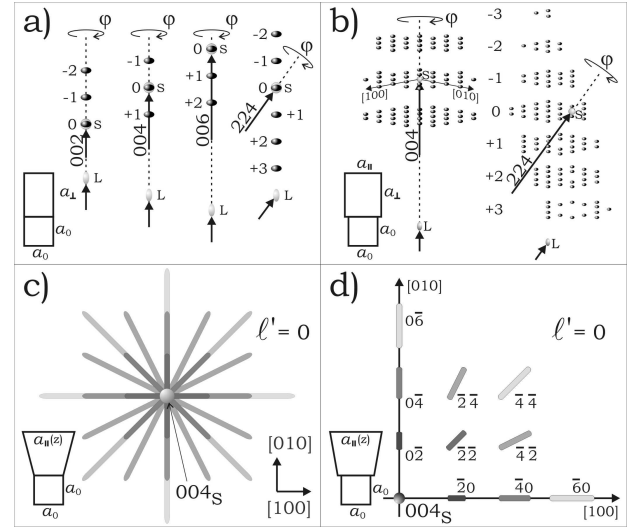}
\caption{Properties of hybrid reciprocal lattice in
epilayer/substrate (001) cubic systems. (a) HRLPs near symmetrical,
002, 004, and 006, and asymmetrical, 224, substrate reflections. In
fully strained layers, $\Delta a^* = \Delta b^* = 0$, equation
(\ref{eq:4}), and hence the HRLPs are aligned along the (001)
growth direction and they are distinguished only by their $\ell'$
index, shown beside of each one of them (dark spots). The HRLP with
$\ell'=0$ coincides with the substrate RLP marked by S. L stands
for the layer RLP. (b) Relaxed layers where $\Delta a^* = \Delta
b^* \neq 0$ provide a three-dimensional hybrid lattice around
either symmetrical, 004, or asymmetrical, 224, primary substrate
reflections. (c), (d) Strain grading along the layer thickness
gives rise to elongated HRLPs towards the substrate one. HRLPs with
index $\ell'=0$ lay on the layer in-plane direction, and (d) they
are distinguishable from the substrate RLP if some relaxation
occurs at the layer/substrate interface, the $h'k'$ indexes are
given. In the insets, $a_{||}$ and $a_{\perp}$ stand for the
in-plane (parallel) and out-plane (perpendicular) unit cell
parameters of the layer, respectively. $a_0$ is the substrate
lattice parameter. In these examples, $a_{\perp} > a_0$, and only
HRLPs in which $|\xi| > 0.008$, equation (\ref{eq:5}), are shown.}
\end{figure}

\begin{figure*}[!]
\includegraphics[width=6.4in]{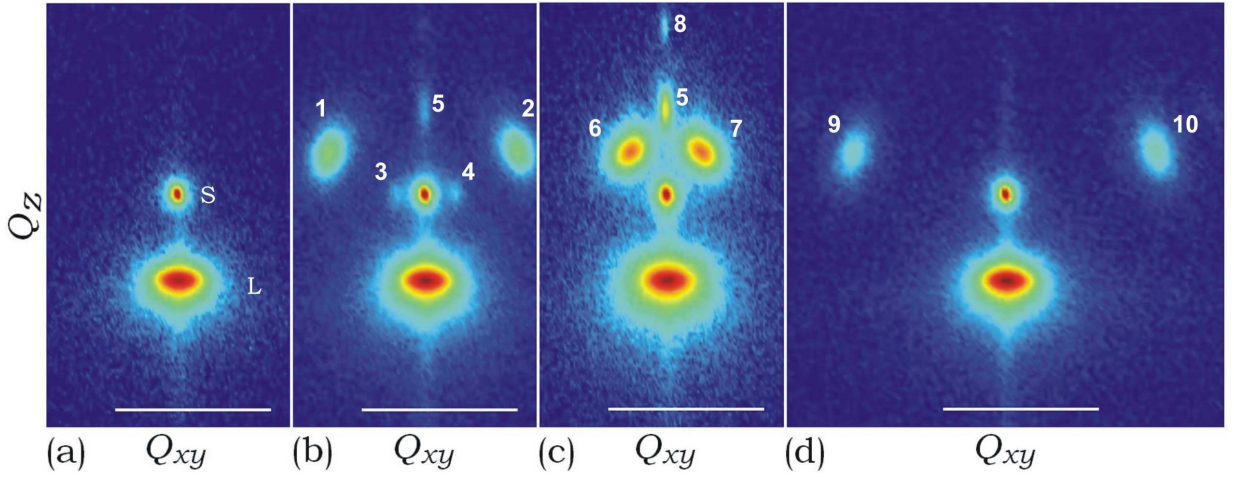}
\caption{Reciprocal-space maps, 002 reflection, ZnSe ($1\mu$m
thick) relaxed layer on GaAs (001), recorded with Cu$K\alpha_1$
radiation at different azimuthal $\varphi$ angles. (a) $\varphi =
0$, only the $002$ substrate (S) and layer (L) RLPs are seen; the
former is at $Q_z = 2/a_0 = 0.35377$\AA$^{-1}$. (b) $\varphi =
27.6^{\circ}$, (c) $\varphi = 30^{\circ}$, and (d) $\varphi =
45^{\circ}$. [110] is the in-plane reference direction for the
$\varphi$ rotation of the sample around the [001] direction.
Visible HRLPs, in (b), (c), and (d), are numbered and identified by
their transversal components $Q_{xy}$ in Table 1. White bars
represent $1.6\times10^{-3}$\AA$^{-1}$. Mesh size:
$2.48\times10^{-5}$\AA$-1$ in $Q_{xy}$ per
$3.27\times10^{-5}$\AA$-1$ in $Q_z$.}
\end{figure*}

\subsubsection{Strain grading}

A unique property of hybrid reflections arises from the layer
reflections since they are always Laue reflections, i.e.
transmitted-diffracted beams where diffraction occurs through the
entire layer thickness. In any other X-ray diffraction technique,
structural analyses of layers are carried out by means of Bragg
reflections: those with reflected-diffracted beams. Therefore, in
principle, HRLPs are equally sensitive to lattice mismatch at both
the top and the bottom of the epilayer. Hence, strain variation
across the layer thickness should generate elongated HRLPs, i.e.
HRL rods, oriented at specific directions on reciprocal-space maps.
Except for the particular cases discussed in Figs. 2(c) and 2(d),
which are sensitive only to the in-plane strain grading, the
general orientations of the rods are affected by the elastic
properties of the layer compound; for example, in the case of (001)
growth on cubic systems with parallel,
$\Delta\varepsilon_{\parallel}$, and perpendicular,
$\Delta\varepsilon_{\perp}$, strain gradings. The $h'k'\ell'$ HRL
rod around a symmetric primary reflection will be oriented at an
angle $\chi_G$ where
\begin{equation}
\tan\chi_G = \frac{\Delta Q_z}{\Delta Q_{xy}} =
\frac{-\ell'}{\sqrt{h'^2+k'^2}\cos(\Delta\varphi)}\frac{\Delta\varepsilon_{\perp}}{\Delta\varepsilon_{\parallel}}
\label{eq:8}
\end{equation}
assuming a tetragonal distortion of the layer unit cell varying as
a function of the strain according to
$a=b=a_L(1+\varepsilon_{\parallel})$ and
$c=a_L(1+\varepsilon_{\perp})$; $a_L$ is the unstressed layer
lattice parameter and, for isotropic materials,
\begin{equation}
\frac{\Delta\varepsilon_{\perp}}{\Delta\varepsilon_{\parallel}}=-\frac{2\nu}{1-\nu}
\label{eq:9}
\end{equation}
where $\nu$ is the Poisson ratio. Being able to predict HRL rod
orientation due to grading is important since there are other
causes for the elongated shapes of the HRLPs, as better explained
below.

\subsubsection{Anisotropic mosaicity}

Besides strain grading and finite layer thickness effects, the
HRLPs may also present elliptical shapes due to mosaicity in the
epilayer. Most diffraction vectors taking part in hybrid
reflections have both in-plane and out-plane components, and
therefore, they are sensitive to the spatial misorientation of the
crystallites, or mosaic blocks, building up the epilayer. In-plane
rotation of the crystallites around the growth direction gives
raise to a mosaicity of width $\eta_{\parallel}$, while crystallite
rotations around in-plane axes give rise to an out-plane mosaic
width $\eta_{\perp}$. It is possible to compute effects of
mosaicity on the shape and orientation of the HRLPs by projecting
the trajectory of the reciprocal vector
\begin{equation}
\Delta\bm{H} =
\eta_{\parallel}(\hat{\bm{z}}\times\hat{\bm{r}})\sin\psi +
\eta_{\perp}[(\hat{\bm{z}}\times\hat{\bm{r}})\times\hat{\bm{r}}]\cos\psi
\label{eq:10}
\end{equation}
on the incidence plane of the reciprocal-space maps as $\psi$
varies through $360^{\circ}$. $\hat{\bm{r}} = \bm{H}/|\bm{H}|$ and
$\bm{H}$ is the layer diffraction vector in
$\bm{P}'=\bm{M}_{S,L}+\bm{N}_{L,S}$, i.e. either $\bm{H}=\bm{M}_L$
or $\bm{H}=\bm{N}_L$. The orientation angle $\chi_M$ is with
respect to the direction of largest projection so that
\begin{equation}
\tan\chi_M = \frac{\Delta Q_z}{\Delta Q_{xy}} =
\frac{\Delta\bm{H}\cdot\hat{\bm{z}}}{\Delta\bm{H}\cdot\hat{\bm{k}}_{xy}}
\label{eq:11}
\end{equation}
when the value of $\Delta Q_z^2+\Delta Q_{xy}^2$ is a maximum.

\begin{table*}[!]
\scriptsize \caption{Hybrid reciprocal lattice points observed in
Figs.~3(b), 3(c), and 3(d), as indicated by HRLP numbers and
$h'k'\ell'$ indexes. $Q_{xy}$ values are obtained by using $\Delta
a^* = \Delta b^* = -5.15\times 10^{-4}$\AA$^{-1}$ in equations
(\ref{eq:4}) and (\ref{eq:6}). $\bm{P}' = \bm{M}_{S,L} +
\bm{N}_{L,S}$ is the hybrid diffraction vector, and $\xi$ is the
cosine director of the secondary beam, equation (\ref{eq:5}). A few
third-order hybrid diffractions are also visible: HRLPs 5, 8, and
9. Exact azimuthal $\varphi_{h'k'\ell'}$ angles where each HRLP is
excited have been calculated as described in Appendix A. The
crystal azimuthal position during data collection is given at the
$\varphi$ column: values with respect to the [110] in-plane
reference direction. $\delta$ is the diffracted beam takeoff angle
as defined in equation (\ref{eq:7}). Elliptical-shape orientation
angles of the HRLPs in the reciprocal-space maps, Fig.~3, are given
by $\chi_E$ (experimental values), while $\chi_G$, and $\chi_M$ are
theoretical values according to two distinct hypotheses, as
explained in the text. Angular values are given in degrees.}
\begin{tabular}{ccccccccccc}
HRLP & $h'k'\ell'$ & $Q_{xy}$~(\AA$^{-1}$) & $\bm{P}'$ & $\xi/|\xi|$ & $\varphi_{h'k'\ell'}$ & $\varphi$ & $\delta$ & $\chi_E$ ($\pm 2$) & $\chi_G$ & $\chi_M$ \\
\hline \\
1 & $\bar{3}\,\bar{1}\,\bar{1}$ & $-9.50\times10^{-4}$ & $313_S+\bar{3}\bar{1}\bar{1}_L$ & $+1$ & 27.7043 & 27.6 & $+0.116$ & $+75$ & $+47.6$ & $+75.2$ \\
2 & $3\,1\,\bar{1}$ & $+9.50\times10^{-4}$ & $31\bar{1}_L+\bar{3}\bar{1}3_S$ & $-1$ & 27.8258 & 27.6 & $-0.116$ & $-75$ & $-47.6$ & $-75.2$ \\
3 & $\bar{2}\,0\,0$ & $-2.92\times10^{-4}$ & $202_S+\bar{2}00_L$ & $+1$ & 28.5022 & 27.6 & $+0.087$ & $+90$ & $0.0$ & $+90.0$ \\
4 & $2\,0\,0$ & $+2.92\times10^{-4}$ & $200_L+\bar{2}02_S$ & $-1$ & 28.5928 & 27.6 & $-0.087$ & $-90$ & $0.0$ & $-90.0$ \\
5 & $0\,0\,\bar{2}$ & \textemdash & $202_S+00\bar{2}_L+\bar{2}02_S$ & $+1$ & 28.5865 & 27.6 & \textemdash & \textemdash & \textemdash & \textemdash \\
6 & $1\,\bar{1}\,\bar{1}$ & $-3.65\times10^{-4}$ & $\bar{1}13_S+1\bar{1}\bar{1}_L$ & $+1$ & 30.1328 & 30.0 & $-0.056$ & $+58$ & $+70.1$ & $+58.5$ \\
7 & $\bar{1}\,1\,\bar{1}$ & $+3.65\times10^{-4}$ & $\bar{1}1\bar{1}_L+1\bar{1}3_S$ & $-1$ & 30.0750 & 30.0 & $+0.056$ & $-58$ & $-70.1$ & $-58.5$ \\
8 & $0\,0\,\bar{2}$ & \textemdash & $\bar{1}1\bar{1}_L+004_S+1\bar{1}\bar{1}_L$ & $-1$ & 30.0369 & 30.0 & \textemdash & \textemdash & \textemdash & \textemdash \\
9 & $0\,0\,\bar{4}$ & \textemdash & $\bar{1}13_S+00\bar{4}_L+1\bar{1}3_S$ & $+1$ & 29.8584 & 30.0 & \textemdash & \textemdash & \textemdash & \textemdash \\
10 & $3\,\bar{3}\,\bar{1}$ & $-15.4\times10^{-4}$ & $\bar{3}33_S+3\bar{3}\bar{1}_L$ & $+1$ & 44.6644 & 45.0 & $-0.136$ & $+80$ & $+33.4$ & $+77.7$ \\
10 & $\bar{3}\,\bar{3}\,\bar{1}$ & $-15.4\times10^{-4}$ & $333_S+\bar{3}\bar{3}\bar{1}_L$ & $+1$ & 45.3356 & 45.0 & $+0.136$ & $+80$ & $+33.4$ & $+77.7$ \\
11 & $\bar{3}\,3\,\bar{1}$ & $+15.4\times10^{-4}$ & $\bar{3}3\bar{1}_L+3\bar{3}3_S$ & $-1$ & 44.5215 & 45.0 & $+0.136$ & $-80$ & $-33.4$ & $-77.7$ \\
11 & $3\,3\,\bar{1}$ & $+15.4\times10^{-4}$ & $33\bar{1}_L+\bar{3}\bar{3}3_S$ & $-1$ & 45.4785 & 45.0 & $-0.136$ & $-80$ & $-33.4$ & $-77.7$ \\
\hline
\end{tabular}
\end{table*}

\section{Results and Discussions}

Occurrence of experimental HRLPs is demonstrated in Fig.~3,
obtained in a sample with a single $1\mu$m thick ZnSe epilayer on
GaAs (001) substrate. Proper identification of the HRLPs is given
in Table 1. The reciprocal-space maps were collected on a Philips
X'Pert-MRD high resolution diffractometer: Cu tube, X-ray mirror,
four-crystal asymmetric 220 Ge monochromator and three-bounce 220
Ge analyzer crystal; nominal spectral width
$\Delta\lambda/\lambda=5\times10^{-5}$; X-ray beam divergences
$0.005^{\circ}$ and $2^{\circ}$ in the incidence plane and in the
axial direction, respectively.

Direct and complete strain analysis of the epilayer is possible
even from reciprocal-space maps of the symmetric 002 GaAs
reflection when HRLPs are excited. The $Q_{xy}$ components of the
HRLPs in Figs. 3(b), 3(c), and 3(d) are well reproduced by using
$\Delta a^* = \Delta b^* = -5.15(12)\times 10^{-4}$\AA$^{-1}$, and
their longitudinal separation (an integer fraction of the
layer-substrate reciprocal-lattice point distance) provides $\Delta
c^* = -4.43(33)\times 10^{-4}$\AA$^{-1}$. Hence, assuming a Poisson
ratio of $\nu = 0.5$, which implies
$\varepsilon_{\perp}/\varepsilon_{\parallel}=-2$, the fully relaxed
lattice parameter of the layer compound is obtained as $a_L =
5.6691(4)$\AA, which is the same as the nominal value of the ZnSe
compound, while the epilayer undergoes an expansive in-plane strain
$\varepsilon_{\parallel}=1.36(54)\times 10^{-4}$.

HRLPs in Fig. 3(d), numbers 10 and 11, have the largest observed
$Q_{xy}$ component, scattering within a theoretical takeoff angle
$\delta = \pm0.14^{\circ}$, equation (\ref{eq:7}), but still
accepted by the analyzer system of the diffractometer. In fact, the
minimum required acceptance of the analyzer system should be, in
general, equal or larger than the axial divergence since the
experimental takeoff angles are increased by an amount
corresponding to the difference between $\varphi_{h'k'\ell'}$ and
$\varphi$ values in Table 1. For instance, HRLPs 3 and 4 would not
appear on the map in Fig.~3(b) if the axial acceptance was below
$1.08^{\circ}$.

HRLPs with null $Q_{xy}$ component, such as those numbered 5, 8,
and 9 in Figs.~3(b) and 3(c), can not be related to second-order
sequences of reflections. $00\ell'$ hybrid reflections do not occur
for partially or fully relaxed epilayers on (001) substrates
because $\Delta a^*$ and $\Delta b^*$ in equation (\ref{eq:4}) have
non-null values. Sequences of third-order, such as
$\bar{1}1\bar{1}_L+004_S+1\bar{1}\bar{1}_L$ and
$\bar{1}13_S+00\bar{4}_L+1\bar{1}3_S$ can explain the $00\bar{2}$
and $00\bar{4}$ type of HRLP even in relaxed layers. Note that when
carrying out longitudinal scans, i.e. $\omega:2\theta$ scans, HRLPs
along the primary reflection axis generate extra features even in
samples with a single epilayer. Moreover, in ordinary rocking
curves, hybrid intensity contributions could be misinterpreted as
due to sublayers or other sort of structural change, such as
composition grading.

Two hypotheses have been verified to explain the elongated shapes
of the experimental HRLPs in Fig.~3. The strain grading hypothesis
should, according to equation (\ref{eq:8}), reduce the absolute
value of $\chi_G$ as the in-plane component of $\bm{P}'$ increases.
The expected values of $\chi_G$, calculated using $\nu=0.5$, are
given in Table 1. Although elongations happen in the same sense as
the observed ones, their behavior as the $h'k'$ indexes increase is
the opposite of that expected. For instance, $\chi_G =
+70.1^{\circ}$, $+47.6^{\circ}$, and $+33.4^{\circ}$, are the
expected values for the HRLPs numbered 6, 1, and 10, while the
observed values are instead $\chi_E = +58(2)^{\circ}$,
$+75(2)^{\circ}$, and $+80(2)^{\circ}$.

\begin{figure*}
\includegraphics[width=3.2in]{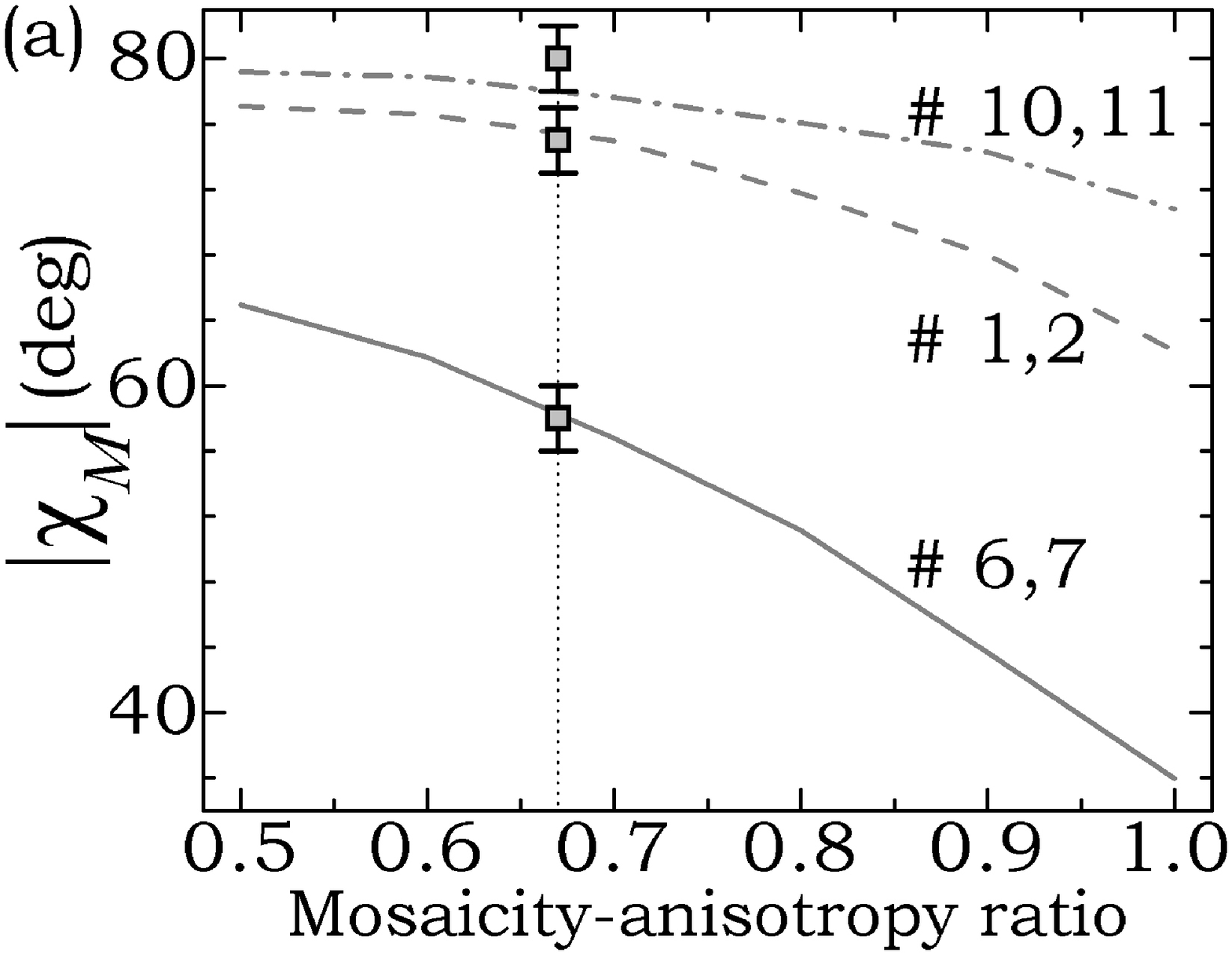}
\includegraphics[width=3.2in]{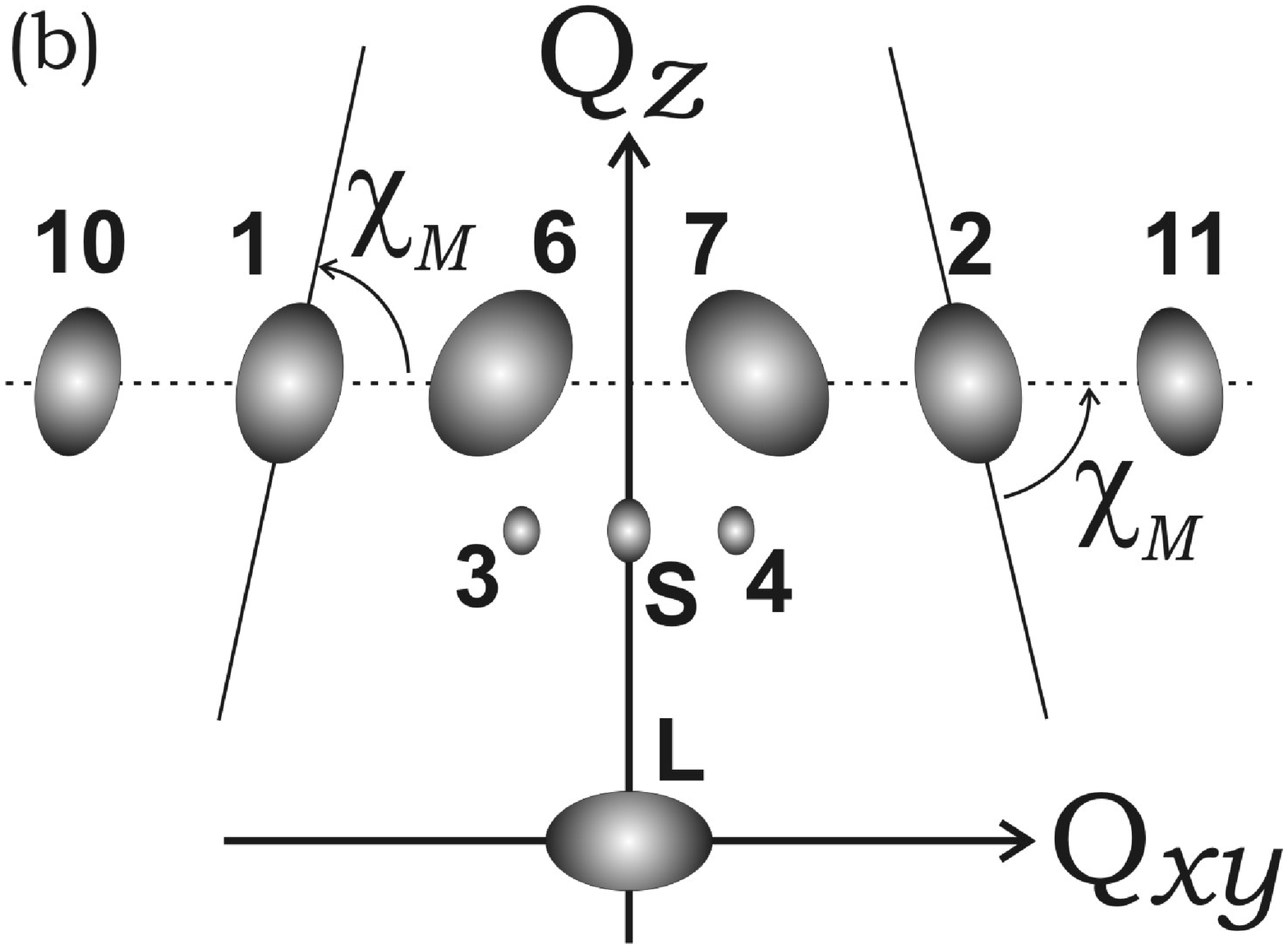}
\caption{(a) Orientation angle $\chi_M$ of the HRLPs 1 and 2
(dashed line), 6 and 7 (solid line), and 10 and 11 (dot-dashed
line), Table 1, as a function of the anisotropy ratio
$\eta_{\parallel}/\eta_{\perp}$ between the in-plane,
$\eta_{\parallel}$, and out-plane, $\eta_{\perp}$, mosaic widths.
Values were computed according to equation (\ref{eq:11}).
Experimental values (squares) from the reciprocal-space maps in
Fig. 3 are also shown. (b) Theoretical shapes and orientations for
67\% of anisotropy in the epilayer mosaicity, numbered and sized
according to Fig.~3. Obtained $\chi_M$ values are compared to the
experimental ones in Table 1.}
\end{figure*}

A successful explanation for the elliptical shapes of the HRLPs
comes from mosaicity in the epilayer. Its 002 reciprocal-lattice
point in Fig.~3 (spot labeled L) has a full width at half-maximum
(FWHM) of $17.5\times10^{-5}$\AA$^{-1}$ along the $Q_{xy}$
coordinate, yielding an out-plane mosaic width of
$\eta_{\perp}~=~0.024^{\circ}$. Fig.~4(a) shows the computed
directions of maximum projection of the $\Delta\bm{H}$ vector onto
the $Q_zQ_{xy}$ maps, described by the angle $\chi_M$, equation
(\ref{eq:11}), as a function of mosaicity-anisotropy ratio
$\eta_{\parallel}/\eta_{\perp}$. A very good match to the
experimental values is obtained for a ratio of $0.67(3)$, as can be
seen by comparing $\chi_M$ and $\chi_E$ in Table 1, and then,
$\eta_{\parallel}~=~0.016(1)^{\circ}$. Projection shapes for this
ratio value is given in Fig.~4(b). In theory (Fig.~4a), hybrid
reflections of indexes $1\bar{1}\bar{1}$ and $\bar{1}1\bar{1}$,
HRLPs 6 and 7, have better sensitivity to the in-plane mosaicity,
and they are used to estimate the $\pm0.03$ error bar of the found
ratio value.

\section{Conclusions}

Hybrid reciprocal lattices for x-ray diffraction in
epilayer/substrate systems exist. Without a systematic description,
as the one provided in this work, accidental excitement of hybrid
reflections in most X-ray diffraction techniques can jeopardize
data analysis. On the other hand, understanding their properties
leads to three-dimensional information of the layer structure even
in symmetrical high-angle diffraction geometries. In the case
analyzed here, from reciprocal-space maps of a single symmetric
reflection we were able to determine parallel and perpendicular
lattice mismatches, the stress state of the epilayer, the absence
of strain grading, and the spatial misorientation of mosaic grains
in the layer.

\appendix
\section{Bragg condition for hybrid reflections}
Let us initially consider a given three-beam diffraction of the
substrate lattice in which $\bm{P}=\bm{M}+\bm{N}$. In the reference
system $(\hat{\bm{x}},\>\hat{\bm{y}},\>\hat{\bm{z}})$ where
$\hat{\bm{z}} = \bm{P}/|\bm{P}|$, $\hat{\bm{x}}$ is an arbitrarily
chosen in-plane direction perpendicular to $\bm{P}$, and
$\hat{\bm{y}} = \hat{\bm{z}}\times\hat{\bm{x}}$, the incident beam
wavevector
\begin{equation}
\bm{k}=-\lambda^{-1}(\cos\omega\cos\varphi\>\hat{\bm{x}}+\cos\omega\sin\varphi\>\hat{\bm{y}}+\sin\omega\>\hat{\bm{z}})
\label{eq:A1}
\end{equation}
fulfills the two Bragg conditions in equations (\ref{eq:1}) and
(\ref{eq:2}) when $\omega = \omega_0$ and $\varphi=\varphi_0$.
Hence, $\omega_0$ is the Bragg angle of the primary reflection,
diffraction vector $\bm{P}$, and $\varphi_0$ can be calculated by
the expression
\begin{equation}
\cos(\varphi_0 - \alpha) = \frac{\lambda|\bm{M}|/2 -
\sin\omega_0\cos\gamma}{\cos\omega_0\sin\gamma} \label{eq:A2}
\end{equation}
given that $\bm{M} =
|\bm{M}|(\sin\gamma\cos\alpha\>\hat{\bm{x}}+\sin\gamma\sin\alpha\>\hat{\bm{y}}+\cos\gamma\>\hat{\bm{z}})$.

The problem is how to calculate the exact incidence,
$\omega=\omega_0+\Delta\omega$, and azimuthal,
$\varphi=\varphi_0+\Delta\varphi$, angles of a hybrid reflection
whose effective diffraction vector is
$\bm{P}'=\bm{M}_{S,L}+\bm{N}_{L,S}$. Since $\bm{P}'$ is not
parallel to the $\hat{\bm{z}}$ axis, equation (\ref{eq:A2}) is no
longer valid for either $\bm{M}_S$ or $\bm{M}_L$ diffraction
vectors of the secondary reflection. To solve this problem we first
wrote (Morelh\~ao, 2002)
\begin{equation}
\bm{k} \simeq \bm{k}_0 +
\frac{\partial\bm{k}}{\partial\omega}\Delta\omega +
\frac{\partial\bm{k}}{\partial\varphi}\Delta\varphi = \bm{k}_0 +
\bm{k}_{\omega}\Delta\omega + \bm{k}_{\varphi}\Delta\varphi
\label{eq:A3}
\end{equation}
and then equations (\ref{eq:2}) and (\ref{eq:3}) were used to build
the system of linear equations
\begin{equation}
\left[\begin{array}{cc}
\bm{k}_{\omega}\cdot\bm{M} & \bm{k}_{\varphi}\cdot\bm{M}\\
\bm{k}_{\omega}\cdot\bm{N} &
\bm{k}_{\varphi}\cdot\bm{N}\end{array}\right]
\left[\begin{array}{c}
\Delta\omega\\
\Delta\varphi\end{array}\right]=- \left[\begin{array}{c}
(\bm{M}/2+\bm{k}_0)\cdot\bm{M}\\
(\bm{N}/2+\bm{M}+\bm{k}_0)\cdot\bm{N}\end{array}\right]
\label{eq:A4}
\end{equation}
where $\bm{M}$ and $\bm{N}$ stand for $\bm{M}_{S,L}$ and
$\bm{N}_{L,S}$, respectively.

The $\varphi_{h'k'\ell'}$ values in Table 1 were calculated by
solving the above equations. For instance, consider the
$\bar{1}\>1\>\bar{1}$ HRLP, line 7 in Table 1. Since the 002 GaAs
reflection is the primary one and the [110] crystallographic
direction was chosen as reference for the $\varphi$ rotation,
$$\hat{\bm{x}} = [1,1,0]/\sqrt{2},\>\hat{\bm{y}} =
[-1,1,0]/\sqrt{2},\>{\rm and}\>\hat{\bm{z}} = [0,0,1].$$ The
corresponding substrate three-beam diffraction occurs at
$$\omega_0=15.8132^{\circ}\>{\rm and}\>\varphi_0=30.0442^{\circ},$$
as obtained from equation (\ref{eq:A2}) since $\gamma=
125.2644^{\circ}$, $\alpha=90^{\circ}$, and
$|\bm{M}|=\sqrt{3}/a_0$. The $\omega_0$ and $\varphi_0$ angles
provide the wavevector $\bm{k}_0$ from equation (\ref{eq:A1}), as
well as $\bm{k}_{\omega}$ and $ \bm{k}_{\varphi}$. By replacing in
equation (\ref{eq:A4})
$$\bm{M}=\bm{M}_L=[-1/a,1/b,-1/c]\;{\rm
and}\;\bm{N}=\bm{N}_S=[1,-1,3]/a_0,$$ with $a=b=5.6699$\AA,
$c=5.6676$\AA, and $a_0=5.6534$\AA, we have
$$\Delta\omega=0.0794^{\circ}\>{\rm
and}\>\Delta\varphi=0.0308^{\circ},$$ which leads to
$\varphi_{h'k'\ell'}=\varphi_0+\Delta\varphi=30.0750^{\circ}$ in
Table 1.

An alternative approach to calculate the $Q_z$ and $Q_{xy}$
coordinates of the HRLPs, equation (\ref{eq:6}), is available after
determining $\Delta\omega$ and $\Delta\varphi$. Since $\bm{k}$ is
known, we also known $\bm{k}' = \bm{P}'+\bm{k}$,
$$\hat{\bm{k}}_{xy}=-(\cos\varphi\>\hat{\bm{x}}+\sin\varphi\>\hat{\bm{y}})
\;{\rm and}\;
\hat{\bm{s}}=(\sin\varphi\>\hat{\bm{x}}-\cos\varphi\>\hat{\bm{y}}).$$
By defining
$$\bm{k}'_{off}=(\bm{k}'\cdot\hat{\bm{s}})\hat{\bm{s}}\;{\rm
and}\;\bm{k}'_{in}=\bm{k}'-\bm{k}'_{off},$$ we have
$$\omega'=\arccos(\hat{\bm{k}}_{xy}\cdot\bm{k}'_{in}/|\bm{k}'_{in}|),$$
and hence
\begin{eqnarray}
Q_z &=& (\sin\omega'+\sin\omega)/\lambda - |\bm{P}| \nonumber \\
Q_{xy} &=& (\cos\omega'-\cos\omega)/\lambda. \nonumber
\end{eqnarray}
This approach also provides an alternative expression,
$$\delta=\arctan(\lambda \bm{k}'\cdot\hat{\bm{s}}),$$ to calculate
the takeoff angle given in the equation (\ref{eq:7}).

\begin{acknowledgments}
Thanks are due to Vaclav Hol\'y (Charles University, Prague, Czech
Republic) for helpful discussions, and to the European Community
program C1MA-CT-2002-4017 (Centre of Excellence CEPHEUS) for
partial support. SLM also acknowledges founding from Brazilian
agencies CNPq (proc. No. 301617/1995-3) and CAPES/GRICES (proc. No.
140/05).
\end{acknowledgments}

\pagebreak

\end{document}